# MorphoNoC: Exploring the Design Space of a Configurable Hybrid NoC using Nanophotonics

Vikram K. Narayana[a,∗], Shuai Sun[a], Abdel-Hameed A. Badawy[b], Volker J. Sorger[a], Tarek El-Ghazawi[a]

[a]*The George Washington University, Department of Electrical and Computer Engineering, 800 22nd St NW, Washington, D.C., 20052*
[b]*New Mexico State University, Klipsch School of Electrical and Computer Engineering, 1125 Frenger Mall, Las Cruces, NM 88003*

## Abstract

As diminishing feature sizes drive down the energy for computations, the power budget for on-chip communication is steadily rising. Furthermore, the increasing number of cores is placing a huge performance burden on the network-on-chip (NoC) infrastructure. While NoCs are designed as regular architectures that allow scaling to hundreds of cores, the lack of a flexible topology gives rise to higher latencies, lower throughput, and increased energy costs. In this paper, we explore MorphoNoCs - scalable, configurable, hybrid NoCs obtained by extending regular electrical networks with configurable nanophotonic links. In order to design MorphoNoCs, we first carry out a detailed study of the design space for Multi-Write Multi-Read (MWMR) nanophotonics links. After identifying optimum design points, we then discuss the router architecture for deploying them in hybrid electronic-photonic NoCs. We then study the design space at the network level, by varying the waveguide lengths and the number of hybrid routers. This affords us to carry out energy-latency trade-offs. For our evaluations, we adopt traces from synthetic benchmarks as well as the NAS Parallel Benchmark suite. Our results indicate that MorphoNoCs can achieve latency improvements of up to 3.0× or energy improvements of up to 1.37× over the base electronic network.

*Keywords:* Network-on-Chip; Nanophotonics; Reconfigurable Networks; Design-space Exploration; Optical Interconnects

## 1. Introduction

Shrinking feature sizes in silicon have contributed to a steady and substantial increase in the number of transistors packed within a single chip. However, single-core performance can only increase as a square root of the available on-chip resources, as captured by Pollack's rule [1]. Additionally, the end of Dennard scaling has prevented any increase in the clock frequency over the past decade, further limiting the performance improvements that can be achieved using a single core [2]. All of this has ushered us into the many-core era to effectively utilize the available transistors and cater to the ever increasing performance demands of embedded and HPC systems.

With a large number of on-chip cores, packet-switched networks-on-chip (NoC) have emerged as a viable solution for serving the communication needs among the cores, as well as for accessing memory. Their structured design allows for scaling to dozens of cores as compared with bus-based designs, while minimizing costs compared to their fully connected counterparts [3]. However, as the number of cores grow into the hundreds, there is a growing gap between the on-chip computational capability (FLOP/s) and the available on-chip bandwidth [4]. Specifically, sharing of the available bandwidth between processors results in an increased overall latency for core-to-core communications. More importantly, structured NoCs exhibit a higher number of hops for distant, communicating nodes, which significantly increases the latency at larger core counts. In addition, the energy consumed for data movement is growing to be a significant fraction of that needed for computations. For instance, even with the older 65 nm technology node, Intel reported that their 80 core TeraFlops processor incurred 28% chip power solely for the routers and links [5].

Applications with traffic patterns involving distant, communicating nodes would thus derive significant performance and energy benefits if the underlying topology can be adapted to their requirements. Nanophotonics is a promising technology for network building blocks, due to their inherently low latency, high throughput, and low dynamic energy requirements [6]. In this paper, we explore the use of nanophotonics to augment electronic NoCs in order to maintain the best of both worlds and enable the required configurability.

A high-level overview of the proposed hybrid NoC is shown in Figure 1, which includes a serpentine waveguide that traverses an electronic mesh NoC. Separate waveguides in the forward and reverse direction are provided in

---

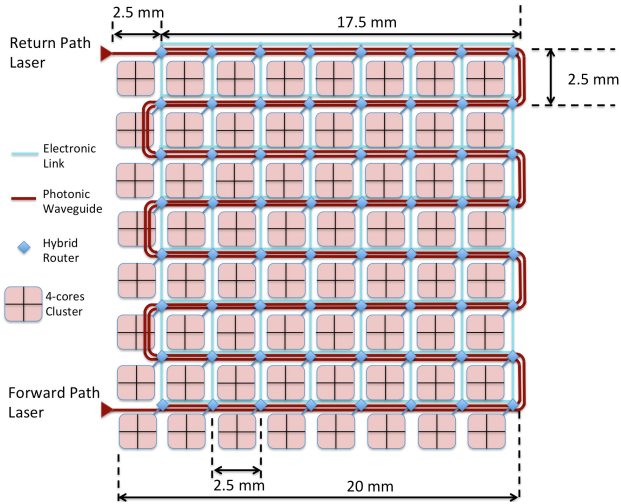

Figure 1: Overview of a basic version of MorphoNoC for an 8x8 mesh.

order to allow bidirectional data transfer. Related work and motivation for our study is outlined in Section 2. We study different NoC versions that involve long as well as medium length nanophotonic waveguides, and collectively term them as MorphoNoCs. Beginning with a multi-write multi-read (MWMR) link as the building block, we optimize the link design parameters to minimize the energy per bit, as detailed in Section 3. We then present a hybrid router design that can host these configurable links, Section 4, and then the overall MorphoNoC architecture at the system level in Section 5. An evaluation of the different flavors of MorphoNoCs is carried out using synthetic benchmarks as well as traces from the NAS Parallel Benchmarks, as summarized in Section 6. Conclusions are finally discussed in Section 7.

## 2. Motivation and Related Work

Several related works have explored the use of photonics in networks-on-chip - a good summary is provided in the literature [7]. Examples include Corona [8], a NoC with MWSR optical loops with token-based aribitration; Flexishare [9], a multi-stage optical crossbar interconnect; a CMP optical bus with dedicated wavelength for each node [10]; LumiNOC, a NoC with multiple optical subnets [11]; an optical NoC with a new structure termed Quartern Topology (QuT) [12]; and ATAC, a hybrid NoC that uses an optical loop for broadcast operations [13]. Purely photonic NoCs typically need a parallel electronic network for establishing the route before transmission, or a separate waveguide for photonic token-based arbitration. Several of these are also concentrated topologies, which means multiple cores are attached to a router node. Configurable channel-based NoCs have also appeared in recent literature [14, 15]. All-optical NoCs that eliminate the need for path arbitration by using wavelength-routed schemes have also been proposed [16, 17]. Nevertheless, arbitration for the ejection channel at the destination node cannot be avoided.

While photonics have the potential for significantly increasing the bandwidth available while reducing the latency, we believe that an all-optical NoC may not be the best option for present day applications; we showed this from a perspective of the performance/cost ratio in our other works [18, 19]. Kennedy and Kodi [14] demonstrated that when real applications are considered, an all-optical NoC only partially utilizes the resources (links). This is true because in real applications, the average injection rate is typically very small ($\approx$0.1) [20]. An alternative strategy, as explored in this work, would instead deploy photonic links only for long-range traffic and for nodes that communicate heavily, and rely on the cheaper and well-understood and easily routable electronics for all other traffic. Furthermore, we expect that the O/E and E/O conversions incur additional clock cycles overhead, thus rendering optical links inferior for short distance traffic between, for instance, neighboring core routers (which takes only 1 clock cycle in electronics). In fact, ATAC [13] adopts a hierarchical strategy of different types of networks, with a base network using an electronic mesh, and augmented with an optical loop. Apart from the difference that their work relies on optics for broadcast-type operations whereas we establish point-to-point links on the same MWMR waveguide, the key differentiating factor between their work and ours is that we study in detail the optimum parameters selection for the optical links. Furthermore, we show that instead of having a long optical loop (serpentine), it could be beneficial to split it up into smaller waveguides to achieve lowered power consumption, thus trading off performance for improved power.

In summary, we believe that our study is complementary to the work in the literature, by not only providing a detailed look at optimizing MWMR links, but also exploring the design space at the network level with trade-offs in performance, energy and resource costs.

Furthermore, with the lack of memory storage in optics (no flip flops or registers or buffers), an all-optical network will require suitable a infrastructure for arbitration and/or routing. For instance, researchers have either used a separate arbitration waveguide [8], used a parallel electronic network for setting up paths [21], or used tokens on the existing optical crossbar [9]. There are overheads associated with arbitration and channel setup before packets begin transfer. On the other hand, we feel that an alternative approach where a base electronic network is utilized while leveraging the photonic advantages for long links, is another useful scenario worth studying in detail. Due to static configuration of the long links (before an application begins), there are no overheads in arbitration and link setup at run-time. Our work also recognizes the fact that electronic NoCs continue to have many benefits in energy and cost, and a hybrid opto-electric NoC appears to be a good option for the near future.

Thus, the contributions of this work are as follows:

- An in-depth energy-efficiency study of MWMR photonic links, demonstrating the selection of the optimum design points under different constraints;

- A design-space exploration of the proposed hybrid NoC at the network level, considering different lengths and number of readers/writers on the waveguides;

- A robust evaluation by separately estimating the static power consumption and the dynamic energy at the network level;

- A realistic exploration by using a low injection rate of 0.1 for design decisions, which brings out the limitations of nanophotonics due to their higher static power.

## 3. Reconfigurable Nanophotonic links

A typical nanophotonic link is composed of a laser source, waveguide, a modulator at the transmitting end, and a photodetector at the receiving end [6]. Photonic interconnects can be point-to-point, single-write multi-read (SWMR), multi-write single-read (MWSR), and multi-write multi-read (MWMR). From the MorphoNoC overview in Figure 1, it is clear that at each hybrid router, we need a mechanism for injecting and receiving data at multiple points from the serpentine waveguides. Thus an MWMR interconnect is well-suited for our purpose. Furthermore, as noted in the literature, the MWMR interconnect offers the greatest flexibility and highest density [22], and is thus adopted for this work.

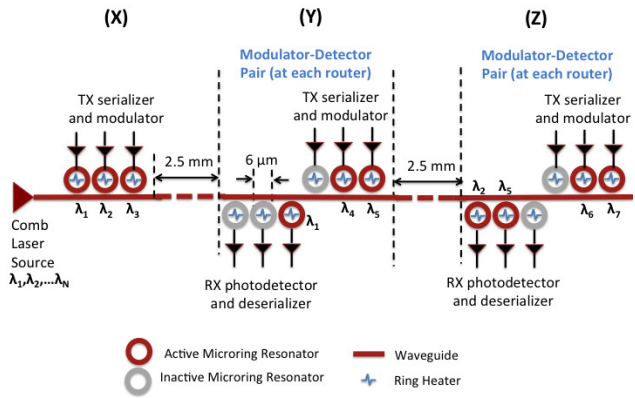

Figure 2: Illustration of MWMR interconnect used in this work. For simplicity, only a single waveguide in one direction is shown.

An illustration of our MWMR interconnect is shown in Figure 2. Each modulator-detector pair as shown is present at every hybrid router. The minimum distance between two routers is 2.5 mm, corresponding to one hop in the base electronic mesh network. The laser source provides a comb of frequencies in order to allow wavelength division multiplexing (WDM) based channels. At each transmitter, parallel data is serialized and then fed to a driver that modulates the microring resonators (MRR). Each MRR is tuned to a particular frequency. By rapidly shifting the resonance of an MRR away from its base frequency and returning it back, light can be selectively retained or removed from the main waveguide to achieve data transmission. At the receiving end, an MRR that is tuned to the same frequency is able to capture the data. The received photonic data is converted back to electrical signals using a photodetector, amplifier and a deserializer. In the example shown in Figure 2, links X→Y, X→Z, and Y→Z are respectively established through $\lambda_1$, $\lambda_2$, and $\lambda_5$. On such an MWMR link, any source-destination pair can establish a link, but not all pairs can simultaneously be connected due to limitations in the number of wavelengths and the number of MRRs available at each node.

The active MRRs at the transmitter and receiver side may experience a drift in their frequency due to thermal variations, and are thus provided with heaters in order to stabilize their resonant frequencies [7]. Ring heating power, also known as trimming power, is considered to be static, and represents a large fraction of the static power dissipation in nanophotonics.

The example shown in Figure 2 is very simplified as it shows only one waveguide, and data from one electrical link is transmitted using a single wavelength. In our MorphoNoC, we use multiple wavelengths in order to encode data from a single electrical link. Furthermore, multiple waveguides are also used in each direction. The maximum number of wavelengths that can be deployed on each waveguide depends on the free spectral range (FSR) for the selected MRRs as well as the bandwidth for each wavelength [23]. For the selected MRR dimensions in our design, the FSR is around 2THz, and we conservatively utilize 512 Gb/s as the available aggregate data rate for each waveguide.

### 3.1. Design Space Parameters

For the base MorphoNoC shown in Figure 1, a number of design space parameters can be varied for the MWMR interconnect in each direction. As explained in Section 4, we assume a 128 Gb/s data rate for each electrical link (hereby referred to as "logical link" when it crosses into the photonic domain). The MWMR design parameters are summarized in Table 1. The parameter *stride* is an integer value indicating the number of router hops spacing between two consecutive modulator-detector (M-D) pairs. A stride value of 1 indicates that an M-D pair is located every 2.5 mm, whereas a stride of 2 is used when the distance is 5 mm, effectively skipping one router along the waveguide. Since each M-D pair incurs signal power losses as well as heating power, different values of stride allow for a trade-off between energy and performance. The maximum number of logical links that can be supported by each waveguide is the ratio of the aggregate data rate (512 Gb/s) and the logical link rate (128 Gb/s). Thus, a maximum of four logical links per waveguide are allowed. The length of the waveguides $L$ is fixed for a given flavor of MorphoNoC, and is a multiple of the per-hop length of 2.5 mm. The value of $E$ is also generally fixed for a given interconnect, chosen among a range of values. The possible values of the MWMR design parameters is shown in Table 1. Note that a logical link on an MWMR uses exactly the same data rate as an electrical link, 128 Gb/s. In other words, if the data rate for each wavelength is 16 Gb/s, a logical link will utilize exactly 8 wavelengths, irrespective of the available number of wavelengths.

Table 1: Design Space Parameters for MWMR Interconnect

| Symbol | Description | Values |
|---|---|---|
| $D_\lambda$ | Data rate per wavelength | {2, 4, 8, 16, 32} Gb/s |
| $N_\lambda$ | # Wavelengths per waveguide | (512 Gb/s / $D_\lambda$) |
| $E$ | Total logical links supported | {4, 8, 16, 32} |
| $W$ | Number of waveguides | {$E/4$, $E/4+1$, …, $N_\lambda$} |
| $L$ | Length of waveguides | {2.5, 5, 7.5, …, 160}mm |
| $S$ | Stride | {1, 2, 4, 8} |

Table 2: Nanophotonic Parameters

| Parameter | Value |
|---|---|
| Technology Node | 11 nm Tri-Gate |
| Waveguide Loss | 100 dB/m |
| Coupler loss | 1 dB |
| Waveguide Bending Loss | ~0 dB |
| Laser Efficiency | 25% |
| Ring Through Loss | 0.01 dB |
| Ring Drop Loss | 1 dB |
| Ring Area | 100 $\mu m^2$ |
| Modulator Insertion Loss | Optimized (0.01-10 dB) |
| Modulator Extinction Ratio | Optimized (0.01-10 dB) |
| Ring Tuning Model | Thermal + Bit Reshuffle |
| Ring Tuning Efficiency | 10 GHz/K |
| Ring Heating Efficiency | 100 K/mW |
| Temperature Range | 280-380 K |
| Injection Rate | 0.1 |
| Target BER | $10^{-15}$ |

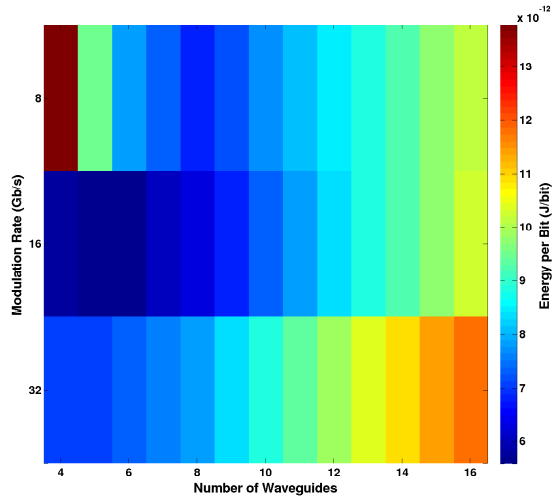

Figure 3: Variation of MWMR interconnect energy per bit with $D_\lambda$ and $W$.

3.2. Energy-efficient Parameter Selection

We modified the DSENT tool in order to model MWMR interconnects. DSENT provides estimates of energy consumption for contemporary nanophotonics, as well as electronic routers and links for technology nodes down to 11nm [20]. We obtained a version of DSENT from the Graphite distribution [24] since it includes the model for an SWMR link, and modified it for our case accordingly. The photonic parameters adopted for our study are summarized in Table 2. The insertion loss and extinction ratio for the ring modulator are automatically optimized by DSENT to achieve the lowest possible modulator and laser power consumption in total. Furthermore, as previously described, MRRs are tuned by heating in order to offset any drifts. In general, the rings can drift across the entire FSR of 2 THz, thereby requiring $\approx$ 2 mW trimming power per ring based on Table 2. With this model, the total power increases with more rings (wavelengths) per waveguide. However, by using the bit reshuffling model in DSENT, any ring that drifts into an adjacent frequency band can be utilized for that band through simple bit reshuffling in hardware [20]. Due to each ring now requiring tuning only across its own frequency band, the total heating power remains constant for a given waveguide irrespective of the number of wavelengths. This information is useful for understanding the simulation results described as follows.

We executed a total of over 700,000 simulations of the modified-DSENT in order to explore the entire design space, covering the parameters in Table 1.

All simulations used an injection rate of 0.1 to compute the energy per bit. We observed that at any given length, there exists an optimum data rate per wavelength, as well as an optimum number of waveguides, in order to achieve the lowest energy per bit. These trends are captured in Figure 3 for $L$=0.07 m, $E$=16, and $S$=1, with 128 Gb/s logical links. In this case, the minimum energy occurs for 6 waveguides and 16 Gb/s per wavelength. These results can be explained by studying the distribution of energy among the different factors. Table 3 shows this data at a fixed number of waveguides, $W$, but with varying $D_\lambda$.

Table 3: Energy Components for $W$=6, $L$=0.07 meter, $E$=16, $S$=1

| $D_\lambda$ (Gb/s) | Energy per Bit (pJ/bit) | | | | |
|---|---|---|---|---|---|
| | Laser | MRR Ht | Lkg | Dyn | Total |
| 8 | 4.11 | 3.38 | 0.10 | 0.21 | 7.81 |
| 16 | 2.06 | 3.38 | 0.10 | 0.18 | **5.71** |
| 32 | 3.50 | 3.38 | 0.09 | 0.33 | 7.29 |

Table 4: Latency Parameters for Photonics [25]

| Parameter | Value |
|---|---|
| Modulator Driver Latency | 9.5 ps |
| Modulator Delay (E-O Conversion) | 14.3 ps |
| Photodetector Delay (O-E Conversion) | 0.2 ps |
| Receiver Amplifier | 4.0 ps |
| Link Propagation | 4.67 ps/mm |

supporting a fixed $E$=16 logical links). As a result, ring losses per waveguide decrease, allowing lower laser power. At the same time, however, the ring heater power is increasing, as it is a constant value per waveguide. Beyond a certain point, the ring heater power dominates, resulting in an increase in the total energy per bit as waveguides are added.

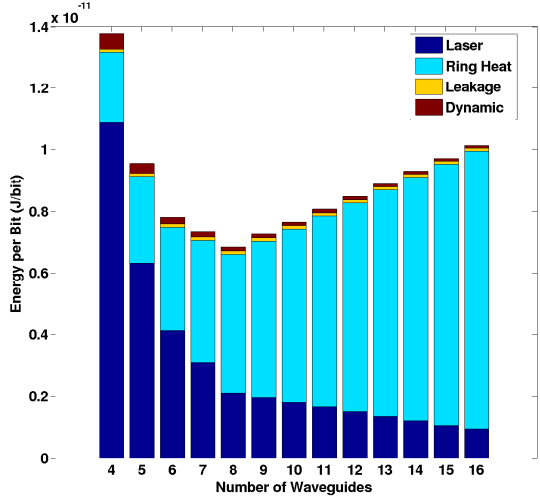

Figure 4: Energy components for $D_\lambda$ = 8 Gb/s, $L$=0.07 m, $E$=16, $S$=1, injection rate=0.1, and 128 Gb/s logical links.

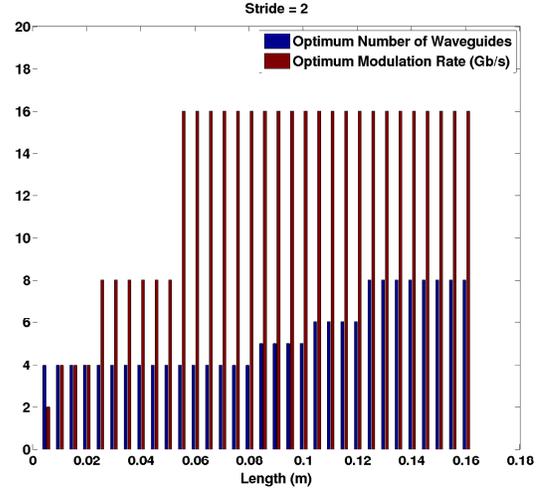

Figure 5: Variation of the optimal $D_\lambda$ and $W$ with length. $E$=16, $S$=2, injection rate=0.1, and 128 Gb/s logical links.

As $D_\lambda$ takes on values of 8, 16, and 32 Gb/s, the number of wavelengths per waveguide changes to 16, 8, and 4, respectively. The ring heating power remains constant, as previously explained. However, as we decrease the number of wavelengths per waveguide from 16 to 8, the number of rings (modulator + detector) is decreasing, yielding lower signal losses on the waveguide. As a result, a lower laser power is sufficient for a good detection at the receiver. However, as we further increase the modulation rate correponding to 32 Gb/s, the laser needs higher power just to sustain the higher data rate. This effect dominates, giving rise to a net increase in energy. The dynamic energy of the associated electronics appears to follow the same trend as the laser power.

Figure 4 shows the energy variation as we change the number of waveguides $W$, keeping the data rate constant at 8 Gb/s per wavelength. At lower $W$, the laser power dominates. When $W$ increases, $N_\lambda$ decreases (as we are

This discussion highlights the need for selecting the optimum data rate and number of waveguides for MWMR interconnects. We expect that at longer lengths $L$, ring losses will dominate due to the increasing number of modulators/detectors along the waveguide. As a result, we can predict that higher data rates will achieve optimal energy, because higher rates will decrease the number of wavelengths and thus the total number of rings along the length of the waveguide. Furthermore, we also predict a larger number of waveguides at longer lengths, in order to spread out the ring losses across waveguides. These trends can be observed in the optimum $D_\lambda$ and $W$ numbers plotted as a function of length $L$ in Figure 5.

*3.3. Latency*

The latency values for our MWMR link are calculated based on the individual latency paramters for the different components, estimated by Chen et al [25]. These parameters are listed in Table 4. For the 160 mm waveguide in Figure 1, the above parameters translate to a 775 ps delay for the longest path, which is well within the electrical clock period of 1 ns. Thus the latency for every photonic link traversal for a single flit is taken as 1 clock cycle in our simulations. This is valid for the entire flit because the data rate of a logical link on the waveguide matches the electrical link data rate.

### 3.4. Data Rates for Long Lengths

There might be concerns that the propagation delay on our long photonic waveguide might restrict the achievable throughput, that is, the data rate per wavelength. However, due to the predictable propagation delay in photonics, data transmission on photonic channels is generally carried out in a wave-pipelined manner - the next bit is transmitted before the previous bit arrives at the destination [26, 27]. Therefore these long waveguides can easily support data rates of 32 Gb/s, the maximum we have adopted here.

### 3.5. Comparison with Regular Electronics

We obtained the energy per bit for our MWMR links at different lengths as specified in Table 1. At each length, for a fixed $E$, and 128 Gb/s logical links, we obtained the optimum $D_\lambda$ and $W$ through an exhaustive search. For comparison with electronics, we modeled a linear chain of routers and links connected back to back for the same distance as the waveguides, and estimated the energy using DSENT. A flit size of 128 bits @ 1 GHz was used, with routers incorporating 4 ports and 4 virtual channels (VC) per port, and a buffer size of 4 flits per VC. The results are shown in Figures 6 and 7 for injection rates of 0.5 and 0.1. We can see that a stride of 1 is inferior to regular electronics at low injection rates, due to high static power in nanophotonics. Nevertheless, photonics can provide remarkable performance benefits due to low latencies at long lengths. In order to reap energy benefits, a stride of 2 or more is required. Furthermore, the length traversed by a flit in an electronic mesh is always smaller than the serpentine path followed by the photonic flits; as a result, stride values > 1 are absolutely needed if we need any energy improvements.

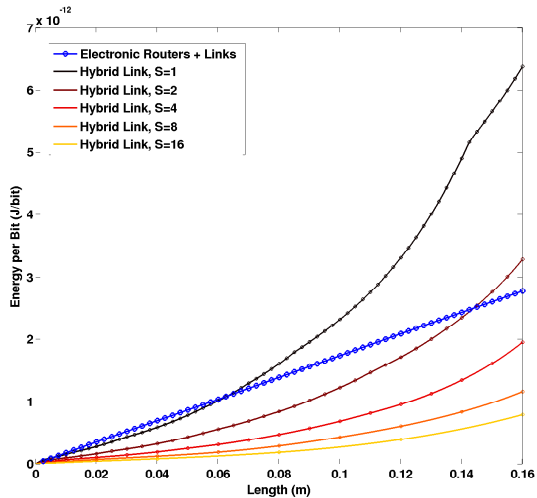

Figure 6: Comparison of Nanophotonic MWMR with Electronics Routers and Links - Injection Rate = 0.5

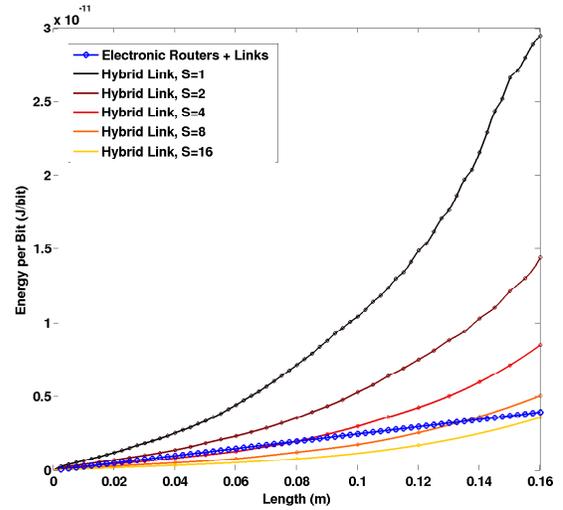

Figure 7: Comparison of Nanophotonic MWMR with Electronics Routers and Links - Injection Rate = 0.1

## 4. Hybrid Router Design

With an optimized MWMR link as the building block, we now present a hybrid router architecture that can allow flits to traverse from the electronic network to the photonic waveguides, and vice versa. The router architecture is shown in Figure 8. The base router from a 2D electronic mesh is extended in order to provide the necessary connections to photonics. Each input port has four virtual channels (VC) - two VCs are reserved for regular traffic from neighboring routers; the third VC is used for flits from neighbors or local processor that need to traverse the photonic pathway. These flits follow the orange path and arbitrate at a smaller *add-on router* before moving onto their intended logical links on the MWMR. The fourth VC is used for serving flits arriving from the photonic links that need to move back into the electrical domain.

The motivations for using two crossbars is two-fold. First, if we assume that the optical serpentine supports 2 logical links in each direction, we would need an additional 4 ports for the router. So the original 5×5 crossbar would now be replaced by a 9×9 crossbar. However, if we split it into two crossbars of size 5×5 and 5×4, the sum of their areas is smaller than that of a single 9×9 crossbar. Second, we believe that using a separate crossbar can give slightly higher performance, explained as follows. When flits arrive at the router input port, they stay temporarily in the virtual channel buffers. The virtual channels associated with each input port of the router will then compete for access to an input port of the crossbar[1]. We want to avoid the optical traffic from competing with regular traffic on the same port of the 9×9 crossbar, given that the optical traffic is guaranteed to take a different output port and thus a different path anyway. By providing a parallel path through a separate crossbar, we avoid this competition. However, the preceding router needs to be aware that the next hop is going to be an optical hop, and assert the appropriate virtual channel request line. As with normal routers, the common data link is used to transmit data to the input of the router.

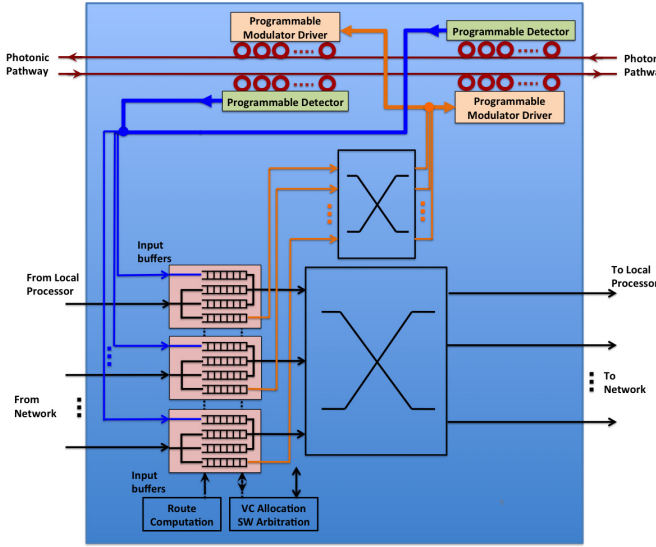

Figure 8: Hybrid Router Architecture.

Table 5: Router design space parameters

| Param. | Description | Value |
|---|---|---|
| $f_e$ | Clock freq. | {4, 2, 1, 0.5} GHz |
| $F$ | Flit size | $128 \times 10^9 / f_e$ = {32,64,128,256} bits |
| $B$ | Buf. depth | 512 / F = {16, 8, 4, 2} flits |

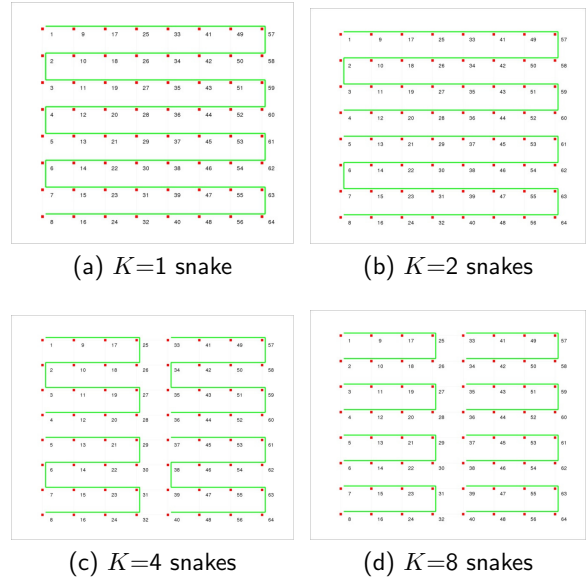

(a) $K$=1 snake   (b) $K$=2 snakes

(c) $K$=4 snakes   (d) $K$=8 snakes

Figure 9: MorphoNoCs for different number of snakes

The modulators and detectors are programmed statically (before the application begins, for instance), to activate the required wavelengths. The programming technique is not studied here, but could most likely make use of out-of-band information from the electronic network path. While any $W$ and $N_\lambda$ values are supported by the architecture, at any given point of time only four outgoing and four incoming photonic links can be programmed in the hybrid router. This is in keeping with the four output ports of the add-on router, as well as four links at the detector output (not separately shown).

---

[1]To avoid competition among virtual channels, some crossbar designs assume that each virtual channel is provided access to a dedicated input port into the crossbar; but this will increase the number of crossbar ports by a factor of n (for n VCs) and thus increase the area significantly.

### 4.1. Energy-efficient Parameter Selection

We used DSENT to model a regular 5-port base router and study different configurations. The data rate is fixed at 128 Gb/s per link. The values of the design parameters are provided in Table 5. The flit sizes are chosen so that the same data rate per link is maintained. Furthermore, to maintain a fair comparison among routers, the total storage is kept constant at 512 bits per port, and the buffer depth per VC is adjusted accordingly. The dynamic energy for the links and router decreases with reducing clock frequency; however, the router leakage, predominantly the buffer leakage, increases despite the constant storage size. This is attributed to larger buffer depth requiring more decoding circuitry. Due to the two opposing factors, the minimum energy per bit point occurs at 128 bits/1 GHz for injection rate = 0.1. We thus chose this configuration for MorphoNoC.

## 5. Putting It All Together: MorphoNoCs

MorphoNoCs can now be designed using the building blocks presented till now. One version is shown in Figure 1. This chip supports 256 cores by utilizing a four-cores cluster at every router node of the network. The dimensions are commensurate with commonly observed numbers for processor cores. The overall design is varied by changing the stride length S for the serpentine MWMR waveguides, as well as the number of sets of serpentine waveguides, or *snakes*, K. MorphoNoCs for different values of K are shown in Figure 9. The design parameters for each snake are optimized as described in Section 3. To make a fair comparison among the different options, we support a constant number of logical links, $E_{tot}$, in the MWMR interconnect, irrespective of the number of snakes K. For instance, with $E_{tot}$=32 at K=8, each of the 8 snakes needs to support only 4 logical links in each direction. The photonic static power dissipation for each flavor of MorphoNoC is shown in Figure 10. The K=1 option is not shown as it is exhibits significantly larger static power and thus does not fit within the scale. For larger number of snakes, the smaller length yields a large reduction in the static power, due to reduced losses from fewer rings and shorter waveguides. To elaborate, the losses affect the laser power exponentially, assuming a fixed responsivity at the receiver end (1.1 A/W). This effect is evidenced from the lower laser power needed for larger number of snakes, as shown in the figure. The different versions allow trade-offs between energy and latency. The hardware resource costs for each of the MorphoNoCs is summarized in Table 6. These are based on the optimum $D_\lambda$, $W$, and $N_\lambda$ derived for each snake as previously outlined. At higher snake counts and stride values, more wavelengths (WL) can be packed into waveguides without significant ring losses, thus giving a lower number of waveguides (WG) and lower modulation rates. In some of the entries in the table, non-integer wavelengths per waveguide is just indicative of the fact that the number of logical links required to be supported by each snake (=$E_{tot}/K$) didn't turn out to be an integer multiple of the number of logical links supported by the optimal waveguide, and thus an extra waveguide with different wavelengths per WG was used.

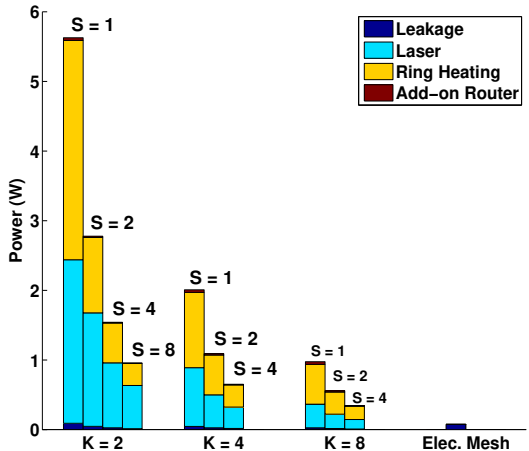

Figure 10: Static photonic power dissipation for different $K$ and $S$

### 5.1. Static Power Comparison with Other Photonic NoCs

A comparison of the static power of MorphoNoCs with other photonic NoCs in the literature is summarized in Table 7. Since the laser power and MRR trimming power are the dominant components (see Figure 10), we focus on these two components. In addition, the table also compares the proposed chip area for these NoCs, as well as the number of MRRs required. For NoCs that have multiple variants, the largest and smallest configurations have been captured in the table. The MorphoNoC variant with only one snake ($K$=1) and unit stride ($S$=1) is inferior in terms of the static power, due to the very long serpentine as well as large MRR aggregated losses that result in a higher demand for laser power. Additionally, for this case, fewer number of wavelengths are used per waveguide (Table 6); as a result, the MRR tuning power cannot be offset significantly by bit shuffling (see Section 3.2), thus yielding 437 $\mu$W trimming power per ring, which is high.

On the other hand, as we increase the number of snakes, the optimum configuration of MorphoNoCs uses a larger number of wavelengths and thus the trimming power per ring reduces to 31 $\mu$W. Moreover, the snakes also become shorter, thus reducing losses and thus lowering the laser power. Additionally, the total number of logical links supported by each snake (=$E_{tot}/K$) reduces as we increase the number of snakes $K$. As a result, each router in each snake needs to support only a fewer number of wavelengths. Overall, the cumulative effect is a reduction in the number of rings for larger $K$, which further drives down the total ring trimming power. Similarly, increasing the stride $S$ reduces the number of hybrid routers, thereby reducing the number of rings as well as losses on the waveguide. Thus, MorphoNoC variants with larger $K$ and larger $S$ are much more power efficient, with the extreme case of $K$=8 and $S$=4 exhibiting itself to be better than the other NoCs. However, this comes at a price, which is a reduced performance due to diminished connectivity, as we shall see in Section 6.

As we just noted, the individual NoC parameters affects the total power consumption. Unfortunately, the different works in the literature do not adopt a uniform set of parameters, and it is thus difficult to make a fair comparison. Table 8 illustrates the different parameters adopted by the different photonic NoCs. The technology node does not significantly affect the static power, because the leakage power from electronic components is small compared with the laser and MRR trimming power. However, the other parameters, namely, the losses, trimming power per ring, and detector efficiency show a large variation among the NoCs. For instance, most NoCs assume ~20 $\mu$W per ring whereas MorphoNoCs shows a range from 31 to 497 $\mu$W per ring, as discussed. Similarly, the MorphoNoCs modulator insertion loss is optimized in the DSENT tool between 0.1-10.0, whereas most other works assume a very small insertion loss. The laser efficiency used by MorphoNoCs is lower (an additional dB of loss). The MRR through loss (or passing-by loss) is 0.01 dB in MorphoNoCs, whereas most of the other works assume 0.001 dB. This will impact the total losses when a large number of rings exist in the system, see Table 7. Thus the loss parameters chosen in MorphoNoCs are very conservative.

Table 6: Resource Counts for Different Versions of MorphoNoCs. These Include Forward and Reverse Paths of the MWMR Links.

| MorphoNoC Type | # WGs | Avg WLs per WG | Length(m) per WG | Add-on Routers | MRRs | $D_\lambda$ (Gb/s) |
|---|---|---|---|---|---|---|
| K=1, S=1 | 64 | 8 | 0.16 | 64 | 65536 | 16 |
| K=1, S=2 | 32 | 16 | 0.16 | 32 | 32768 | 16 |
| K=1, S=4 | 32 | 16 | 0.16 | 16 | 16384 | 16 |
| K=1, S=8 | 32 | 32 | 0.16 | 8 | 16384 | 8 |
| K=2, S=1 | 24 | 21.33 | 0.08 | 64 | 32768 | 16 |
| K=2, S=2 | 16 | 32 | 0.08 | 32 | 16384 | 16 |
| K=2, S=4 | 16 | 64 | 0.08 | 16 | 16384 | 8 |
| K=2, S=8 | 16 | 64 | 0.08 | 8 | 8192 | 8 |
| K=4, S=1 | 16 | 32 | 0.04 | 64 | 16384 | 16 |
| K=4, S=2 | 16 | 64 | 0.04 | 32 | 16384 | 8 |
| K=4, S=4 | 16 | 128 | 0.04 | 16 | 16384 | 4 |
| K=8, S=1 | 16 | 64 | 0.02 | 64 | 16384 | 8 |
| K=8, S=2 | 16 | 128 | 0.02 | 32 | 16384 | 4 |
| K=8, S=4 | 16 | 128 | 0.02 | 16 | 8192 | 4 |

Relying on the DSENT tool, the detector sensitivity in MorphoNoCs is computed based on the chosen (optimized) insertion loss and extinction ratio of the modulator; it thus varies depending on the configuration chosen. The sensitivity varies from -22 dBm to -28 dBm, which indicates that sensing capability is very high (which will reduce the laser power demand). The fixed quantity is the photodetector responsivity, which is set at 1.1 A/W. On the other hand, the other NoCs use detectors with poorer sensitivity, around -20 dBm. Overall, it is difficult to estimate whether the net effect of conservative loss parameters in MorphoNoCs but a superior detector sensitivity will cancel out, as compared with other NoCs. It suffices to note that MorphoNoCs is very competitive in terms of the power consumption, especially for larger $K$ and $S$ values.

### 5.2. Photonic Links Selection

For a given version of MorphoNoC, a simple configuration algorithm is followed in order to select the logical links to be configured within the snakes. The selection is based on the observed traffic pattern for a given application. The algorithm is listed in Algorithm 11, and is self-explanatory.

## 6. Experimental Evaluation

In order to evaluate our MorphoNoCs, we use traces from synthetic benchmarks as well as benchmark suites that run on parallel HPC platforms. We use BookSim 2.0 simulator [30] in trace mode to obtain latency estimates. For energy estimates, we obtain the dynamic energy consumption per flit from our modified DSENT, and use it to compute the total dynamic energy based on the communication volume and the network paths taken by the flits. Static power consumption has already been provided in Section 5. We adopt the 8x8 networks previously described, that are capable of supporting 256 cores. All traffic traces are based on 64-node benchmarks, as the network has 64 routers.

### 6.1. Synthetic benchmarks

Synthetic traffic patterns are obtained from PacketGenie [31]. Specifically, five of the test vectors from the developer's website are directly used for our network simulations. These include the frequently communicating pair (FCP) traces: *FCP Side* and *FCP Center*; as well as the many-to-few-to-many (MFM) patterns that include models of memory at different parts of the chip: *MFM Side*, *MFM Center*, and *MFM Corner*. Simulation results are shown in Figure 12 and Figure 13. For small stride and number of snakes, as expected the latency improvements are better. For K=1 snake, the *FCP Center* and *MFM Center* benchmarks show respective latency improvements of 3.0× and 2.53× over the base network. However, the dynamic energy per bit is inferior, with 2.27× and 2.28× higher dynamic energy over the base network. The situation reverses for larger values of $K$ and $S$. These same applications at K=8 and S=4 have 0.81× and 1.07× latency improvement respectively, and an energy improvement of 1.13× and 1.18× respectively. All the other K and S values provide a range of trade-offs between these two extremes. It is worth noting that certain communication patterns which have medium/short distance communication show good latency results at larger K values, e.g. the *MFM Side* and *MFM Corner*. Such apps are able to derive simultaneous latency and energy benefits, unlike long-distance traffic apps such as *FCP Side* and *FCP Center*.

Table 7: Static Power and Resource Comparison of Photonic NoCs.

| Configuration | Cores | Chip Area | Rings | Laser (W) | Trim (W) |
|---|---|---|---|---|---|
| **Corona** (MWSR Serpentine, 4N cores) [8, 28] | | | | | |
| N=64 | 256 | $625 mm^2$ | 1,056,000 | 2.6 | 26 |
| **Flexishare** (N = kC cores, Multistage Hybrid Opto-Electric, Radix=k, #Channels=M) [9] | | | | | |
| N=64, M=16, K=32, C=2 | 64 | - | 550,000 | 5.8 | 11.9 |
| N=64, M=2, K=16, C=4 | 64 | - | 33,000 | 0.5 | 0.6 |
| **Optical Bus** (4N cores, Optical bus, K wavelengths per node) [10] | | | | | |
| N=8, K=1 | 32 | 400 $mm^2$ | Ignored | 6.35 | 0 |
| N=4, K=1 | 16 | 400 $mm^2$ | Ignored | 0.79 | 0 |
| **LumiNoC** (N cores, Optical subnets using L layers) [11, 29] | | | | | |
| N=64, L=4 | 64 | 400 $mm^2$ | 65,000 | 1.54 | 1.31 |
| N=64, L=1 | 64 | 400 $mm^2$ | 16,000 | 0.35 | 0.33 |
| **QuT** (N cores, Quartern topology) [12] | | | | | |
| N=64 | 64 | 225 $mm^2$ | 45,056 | 0.7 | 0.99 |
| N=128 | 128 | 225 $mm^2$ | 172,000 | 6.52 | 3.79 |
| **CW/CCW** (4N cores, Clockwise and Counterclockwise waveguides) [14] | | | | | |
| N=16 | 64 | 400 $mm^2$ | 104,000 | 2.5 | 2.6 |
| **MorphoNoCs** (4N cores, MWMR Serpentine Opto-Electric, K snakes, S stride) [this work] | | | | | |
| N=64, K=1, S=1 | 256 | 400 $mm^2$ | 65,536 | 6.75 | 16.54 |
| N=64, K=8, S=4 | 256 | 400 $mm^2$ | 8,192 | 0.14 | 0.19 |

Table 8: Comparison of Different Parameters Used by Photonic NoCs.

| Type | Parameter | **Corona** | **Flexishare** | **Optical Bus** | **LumiNoC** | **QuT** | **CW/CCW** | **MorphoNoCs** |
|---|---|---|---|---|---|---|---|---|
| Technology | Process Node | 16 nm | 22 nm | 32 nm | 22 nm | 22 nm | - | 11 nm |
| Efficiency | Laser Efficiency (dB) | 5 | 5 | 0 | 5 | 5 | 5 | 6 |
| | Detector Sensitivity (dBm) | -27.2 | -20 | -15.64 | -20 | -17 | -20 | -22 to -28 |
| Loss | Coupler (dB) | 1 | 1 | 3 | 1 | 1 | 0 | 0 |
| | Splitter (dB) | 0.1 | 0.2 | 0.2 | 0.2 | 0.1 | 0 | 0 |
| | E-O or Diode Loss (dB) | 0 | 1 | 0 | 1 | 0 | 0 | 1 |
| | Modulator Insertion (dB) | 0 | 0.001 | 1 | 0.001 | 0.01 | 0.0001 | 0.01 to 10 |
| | Propagation (dB/cm) | 0.3 | 1 | 1.3 | 1 | 1 | 1 | 1 |
| | WG Bending (dB) | 0 | 0 | 0.5 | 0 | 0.005 | 0 | 0 |
| | WG Crossing (dB) | 0 | 0 | 0 | 0 | 0.12 | 0.05 | 0 |
| | Ring Through (dB) | 0.0016 | 0.001 | 0 | 0.001 | 0.01 | 0.0001 | 0.01 |
| | Ring Drop (dB) | 0 | 1.5 | 0 | 1.5 | 0.5 | 1 | 1 |
| | Interlayer Coupling (dB) | 0 | 0 | 1 | 0 | 0 | 0 | 0 |
| | Detector Loss (dB) | 3 | 0.1 | 0.97 | 0.1 | 0 | 0 | 1 |
| Trimming | Power per Ring ($\mu$W) | 23.6 | 20 | 0 | 20 | 20 | 26 | 31 to 497 |

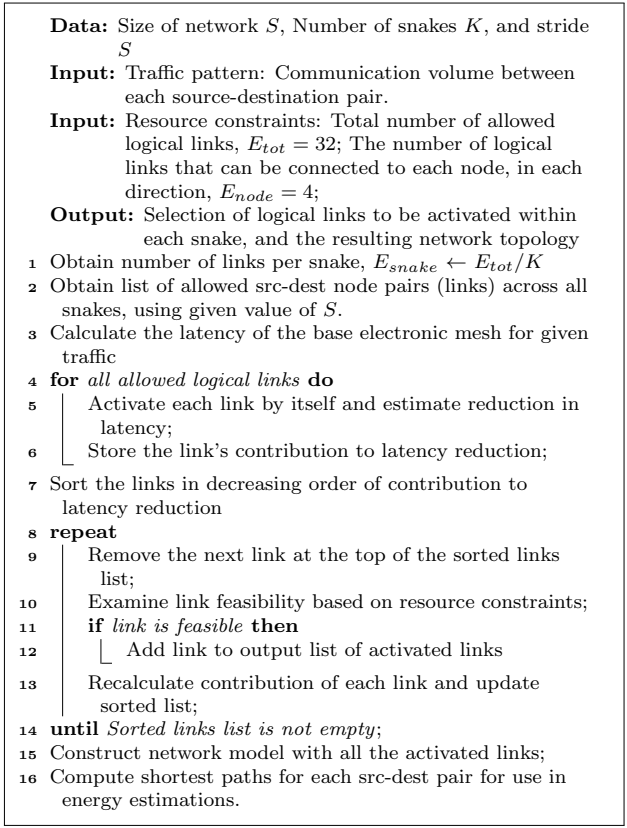

Figure 11: Algorithm for Selecting the Set of Links to be Configured

**Data:** Size of network $S$, Number of snakes $K$, and stride $S$
**Input:** Traffic pattern: Communication volume between each source-destination pair.
**Input:** Resource constraints: Total number of allowed logical links, $E_{tot} = 32$; The number of logical links that can be connected to each node, in each direction, $E_{node} = 4$;
**Output:** Selection of logical links to be activated within each snake, and the resulting network topology

1. Obtain number of links per snake, $E_{snake} \leftarrow E_{tot}/K$
2. Obtain list of allowed src-dest node pairs (links) across all snakes, using given value of $S$.
3. Calculate the latency of the base electronic mesh for given traffic
4. **for** *all allowed logical links* **do**
5.    Activate each link by itself and estimate reduction in latency;
6.    Store the link's contribution to latency reduction;
7. Sort the links in decreasing order of contribution to latency reduction
8. **repeat**
9.    Remove the next link at the top of the sorted links list;
10.    Examine link feasibility based on resource constraints;
11.    **if** *link is feasible* **then**
12.       Add link to output list of activated links
13.    Recalculate contribution of each link and update sorted list;
14. **until** *Sorted links list is not empty*;
15. Construct network model with all the activated links;
16. Compute shortest paths for each src-dest pair for use in energy estimations.

### 6.2. NAS parallel benchmarks

In order to further evaluate our networks with realistic scenarios, we used the NAS Parallel Benchmarks (NPB) [32]. Class A workloads were used for the following kernels - CG, MG, FT, LU, and EP. These benchmarks were executed on an in-house cluster and traffic traces obtained using MPICL. The traces were then converted into BookSim-compatible traces. The simulation results are shown in Figure 14 and Figure 15. The latency results obtained from these benchmarks are more promising. For all types of snakes, there is always an improvement in the average latency. The best latency improvements are observed for MG and FT ($2.37\times$ and $2.60\times$) for K=1, because both these apps have long-range communications. FT also exhibits all-to-all communication and thus shows marked improvement. Curiously, though, the best latency results for both are achieved at stride S=4. Further investigation is needed to understand this result. The LU benchmark is almost completely comprised of 1 hop communications, and thus doesn't see much improvements. The EP (embarrassingly parallel) benchmark has very little communication, but the little data transfer between the root node and all other nodes accounts for the reported latency improvement.

Energy trends are also similar to the synthetic benchmarks, except for LU which shows no energy benefits due to 1 hop traffic. EP consumes very low energy due to the almost non-existent traffic. FT shows the highest energy improvement of $1.37\times$ at K=4 and S=4. At this point, it also has a $2.23\times$ latency improvement over the base network. Improvements in both is attributed to the all-to-all traffic pattern.

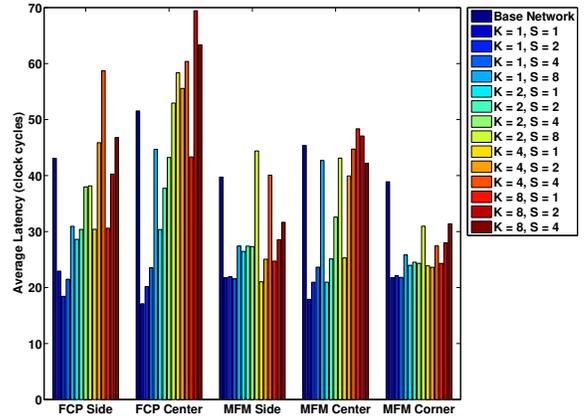

Figure 12: Latency for Different MorphoNoCs - Synthetic Benchmarks

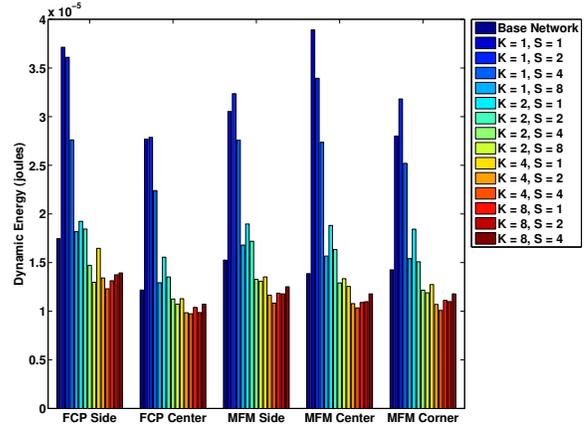

Figure 13: Dynamic Energy for Different MorphoNoCs - Synthetic Benchmarks

### 7. Conclusions

In this paper, we explored an interesting class of hybrid NoCs, that we term as MorphoNoCs. As the number of cores grow, there is an increasing need for high-performance interconnect to cater to long-distance communication. We believe that hybrid opto-electric NoCs that leverage the advances in electronic NoCs and routing techniques, while adopting nanophotonics for long links, would be a possible path forward before we migrate to fully optical NoCs. Our investigations revealed that long, serpentine waveguides, while providing the best performance, can expend considerable power. Different variants of MorphoNoCs were then presented, enabling tradeoffs between performance and power. By carrying out an exhaustive design-space exploration for the individual MWMR links used in our NoCs, we ensured that each variant of MorphoNoC was energy efficient. Moreover, the MMWR exploration also demonstrated the need for choosing suitable parameters for long optical links. Overall, results using synthetic benchmarks as well as the NAS Parallel Benchmarks were promising, indicating latency improvements of up to $3.0\times$ or energy improvements of up to $1.37\times$ over the base electronic network.

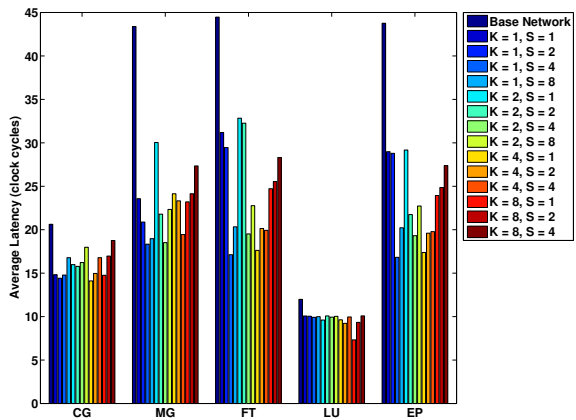

Figure 14: Latency for Different MorphoNoCs - NPB Benchmarks

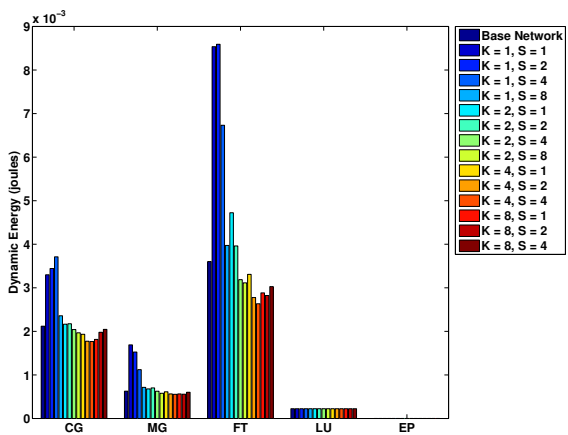

Figure 15: Dynamic Energy for Different MorphoNoCs - NPB Benchmarks

## 8. Acknowledgment

This work was partially supported by the Air Force Office of Scientific Research (AFOSR) under Award FA9550-15-1-0447. The authors also thank the anonymous reviewers for their comments, which helped to improve the quality of the paper.

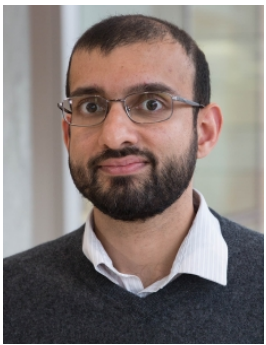
**Vikram K. Narayana** received the received the B.E. degree in electronics and communication engineering from Mysore University, India in 2000, and the Ph.D. degree from the Indian Institute of Technology Madras, India, in 20062007. He is currently an Assistant Research Professor in the Department of Electrical and Computer Engineering at The George Washington University. In the past, he has worked for Wipro Technologies, Siemens Corporate Technology, and STMicroElectronics in Bangalore, India. His research interests include computer architecture, reconfigurable and GPU computing, high-performance computing, photonic network-on-chips and optical computing. He is a Senior Member of the IEEE and IEEE Computer Society.

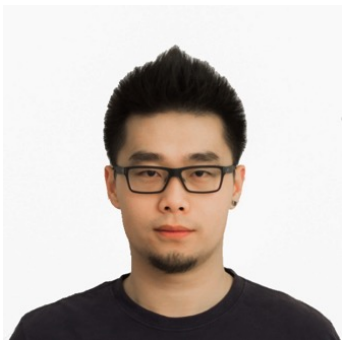
**Shuai Sun** was born in Henan, China in 1990. He received the B.S. degree from North China Electric Power University (Beijing) in 2012 major in Automation and M.S. degree from George Washington Uni- versity in 2014 major in Electrical Engineering. He started his Ph.D. program in George Washington University since 2015 as a research assistant in the Nanophotonic Lab led by Prof. Volker J. Sorger. His research area includes optical NoCs in the levels of nano-devices and network architectures, photonic-plasmonic hybrid interconnects, optical computing, optical neural networks and nanophotonic devices. He is now working on a project which brings together expertise in nano-plasmonics, optoelectronic, architecture, reconfigurable technologies and HPC to address next generation network on chip (NoC) for future optical computing and sponsored by AFOSR.

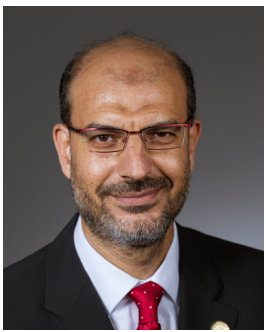
**Abdel-Hameed Badawy** is a tenure-track assistant professor in the Klipsch School of Electrical and Computer Engineering at the New Mexico State University and also held a lead research scientist position at the High Performance Computing Laboratory (HPCL) at the George Washington University. He received his Ph.D. and M.Sc. both from the University of Maryland, College Park, in Computer Engineering. His research interests include locality optimizations, interactions of computer architectures and compilers, high performance computing, machine intelligence techniques and their applications to computer architecture, and Green Computing. He has published in ACM/IEEE conferences and Journals and served on many conference technical program committees. He is a Senior Member of the IEEE, IEEE Computer Society, a Professional member of the ACM, and a deans member of the ASEE. He served as the vice chair of the Arkansas River Valley IEEE section in 2014 and 2016.

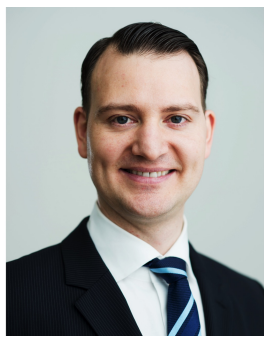
**Volker J. Sorger** is an assistant professor in the Department of Electrical and Computer Engineering, and the director of the Orthogonal Physics Enabled Nanophotonics (OPEN) Labs at the George Washington University. He received his PhD from the University of California Berkeley. His research areas include opto-electronic devices, optical information processing, and internet-of-things technologies. Dr. Sorger received multiple awards such as the Air Force Office of Scientific Research young investigator award, Outstanding Young Researcher Award from GWU, MRS Graduate Gold award, and Intel Fellowship. Dr. Sorger is the executive chair for the technical groups of OSA. He serves at the board of meetings for both OSA and SPIE. He is the editor-in-chief for the journal Nanophotonics, and member of IEEE, OSA, SPIE, and MRS. Lastly he is the founder of the Materials for Nanophotonics subcommittee at the Integrated Photonics Research (IPR) topical meeting, and served on a task force of the National Photonics Initiative (NPI).

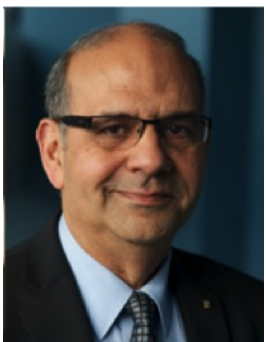
**Tarek El-Ghazawi** is a Professor in the Department of Electrical and Computer Engineering at The George Washington University, where he leads the university-wide Strategic Program in High- Performance Computing. He is the founding director of The GW Institute for Massively Parallel Applications and Computing Technologies (IMPACT). His research interests include high-performance computing, parallel computer architectures, high-performance I/O, reconfigurable computing, experimental performance evaluations, computer vision, and remote sensing. He has published over 200 refereed research papers and book chapters in these areas and his research has been supported by DoD/DARPA, NASA, NSF, and also industry, including IBM and SGI. He is the first author of the book UPC: Distributed Shared Memory Programming, which has the first formal specification of the UPC language used in high-performance computing. Dr. El-Ghazawi is a member of the ACM and the Phi Kappa Phi National Honor Society; he was also a U.S. Fulbright Scholar, a recipient of the Alexander Schwarzkopf Prize for Technological Innovations and a recipient of the Alexander von Humboldt research award from the Humboldt Foundation in Germany. He is a fellow of the IEEE.